\documentclass{article}
\usepackage{amssymb}
\usepackage{graphicx}
\newtheorem{theorem}{Theorem}

\newtheorem{definition}[theorem]{Definition}
\newtheorem{example}[theorem]{Example}

\newtheorem{lemma}[theorem]{Lemma}

\newtheorem{proposition}[theorem]{Proposition}

\def\>{\right>}
\def\<{\left<}

\newcommand{\1}{1\hspace*{-0.255em}{\rm I}}
\begin{document}

\title{Recognition and Teleportation}
\author{Karl-Heinz Fichtner\\
Friedrich-Schiller-Unversit\"at Jena,\\
Institut f\"ur Angewandte Mathematik,\\
 07740 Jena, Germany,\\ 
{\em E-Mail: fichtner@mathematik.uni-jena.de}
\and 
Wolfgang Freudenberg\\
Brandenburgische Technische Universit\"at Cottbus,\\
 Institut f\"ur Mathematik,  PF 101344, 
03013 Cottbus, Germany,\\ 
{\em E-Mail: freudenberg@math.tu-cottbus.de}
\and  Masanori Ohya\\
Science University of Tokyo,\\
Department of Information Science,\\
 Noda City, Chiba 278-8510, Japan, \\
{\em E-Mail: ohya@is.noda.sut.ac.jp}}
\maketitle

\section{Introduction}

We study a possible  function of brain, in particular, we try to describe several aspects of the
process of recognition. In order to understand the fundamental parts of the
recognition process, the quantum teleportation scheme\cite
{BBCJPW,AO2,FO1,FO2} seems to be useful. We consider a channel expression of the
teleportation process that serves for a simplified description of the
recognition process in brain.

It is the processing speed that we take as a particular character of the
brain, so that the high speed of processing in the brain is here supposed to
come from the coherent effects of substances in the brain  like quantum
computer, as was pointed out by Penrose. Having this in our mind, we propose
a model of brain describing its function as follows:

The brain system $\emph{BS}$ =$\frak{X}$ is supposed to be described by a
triple ( $B(\mathcal{H)}$, $\mathcal{S}(\mathcal{H)}$, $\Lambda ^{\ast }(G)$
) on a certain Hilbert space $\mathcal{H}$ where $B(\mathcal{H)}$is the set
of all bounded operators on $\mathcal{H}$, $\mathcal{S}(\mathcal{H)}$ is the
set of all density operators and $\Lambda ^{\ast }(G)$ is a channel giving a
state change with a group $G$.

Further we assume the following:

(1) $\emph{BS}$ is described by a quantum state and the brain itself is
divided into several parts, each of which corresponds to a Hilbert space 
so that $\mathcal{H}$ =$\oplus _{k}\mathcal{H}_{k}$ and $\varphi =\oplus _{k}%
\mathcal{\varphi }_{k},$ $\mathcal{\varphi }_{k}\in \frak{S}(\mathcal{H}_{k}%
\mathcal{)}$. However, in this paper we simply assume that the brain is in
one Hilbert space $\mathcal{H}$ because we only consider the basic mechanism
of recognition.

(2)  The function (action) of the brain is described by a channel $\Lambda ^{\ast }$=$%
\oplus _{k}\Lambda _{k}^{\ast }$. Here as in (1) we take only one channel $%
\Lambda ^{\ast }.$

(3) $\emph{BS}$ is composed of two parts; information processing part ''$P$%
'' and others ''$O$'' (consciousness, memory, recognition) so that $\frak{X=}%
\frak{X}_{P}\otimes \frak{X}_{O}$, $\mathcal{H}$ =$\mathcal{H}_{P}\mathcal{%
\otimes H}_{O}$.

Thus in our model the whole brain may be considered as a parallel quantum
computer\cite{O2}, but we here explain the function of the brain as a quantum
computer, more precisely, a  quantum communication process with entanglements
like in a quantum teleportation process. We will explain the mathematical
structure of our model.

Let  $s=\left\{ s^{_{1}},s^{_{2}},\cdots ,s^{n}\right\} $ be a given (input)
signal (perception) and  $\overline{s}=\left\{ \overline{s}^{_{1}},%
\overline{s}^{_{2}},\cdots ,\overline{s}^{n}\right\} $ the output signal.
After the signal $s$ enters the brain, each element $s^{j}$ of $s$ is coded
into a proper quantum state $\rho ^{_{j}}\in \mathcal{S}\left( \mathcal{H}%
_{P}\right),$ so that the state corresponding to the signal $s$ is $\rho =\otimes _{j}\rho ^{_{j}}.$ 
This state may be regarded as a state processed
by the brain and it is coupled to a state $\rho _{O}$ stored  as a memory
(pre-conciousness) in brain. The processing in the brain is expressed by a properly chosen
quantum channel $\Lambda ^{\ast }$ (or $\Lambda _{P}^{\ast }\otimes $ $%
\Lambda _{O}^{\ast })$. The channel is determined by the form of the
network of neurons and some other biochemical actions, and its function is
like a (quantum) gate in quantum computer\cite{O1,OV1}. The outcome state $%
\overline{\rho }$ contacts with an operator $F$ describing the work as noema
of consciousness (Husserl's noema), after the contact a certain reduction of
state is occured, which may correspond to the noesis (Husserl's) of
consciousness. A part of the reduced state is stored in brain as a memory.
The scheme of our model is represented in the following figure.

\begin{center}
\includegraphics[width=10.0cm,height=10.0cm]{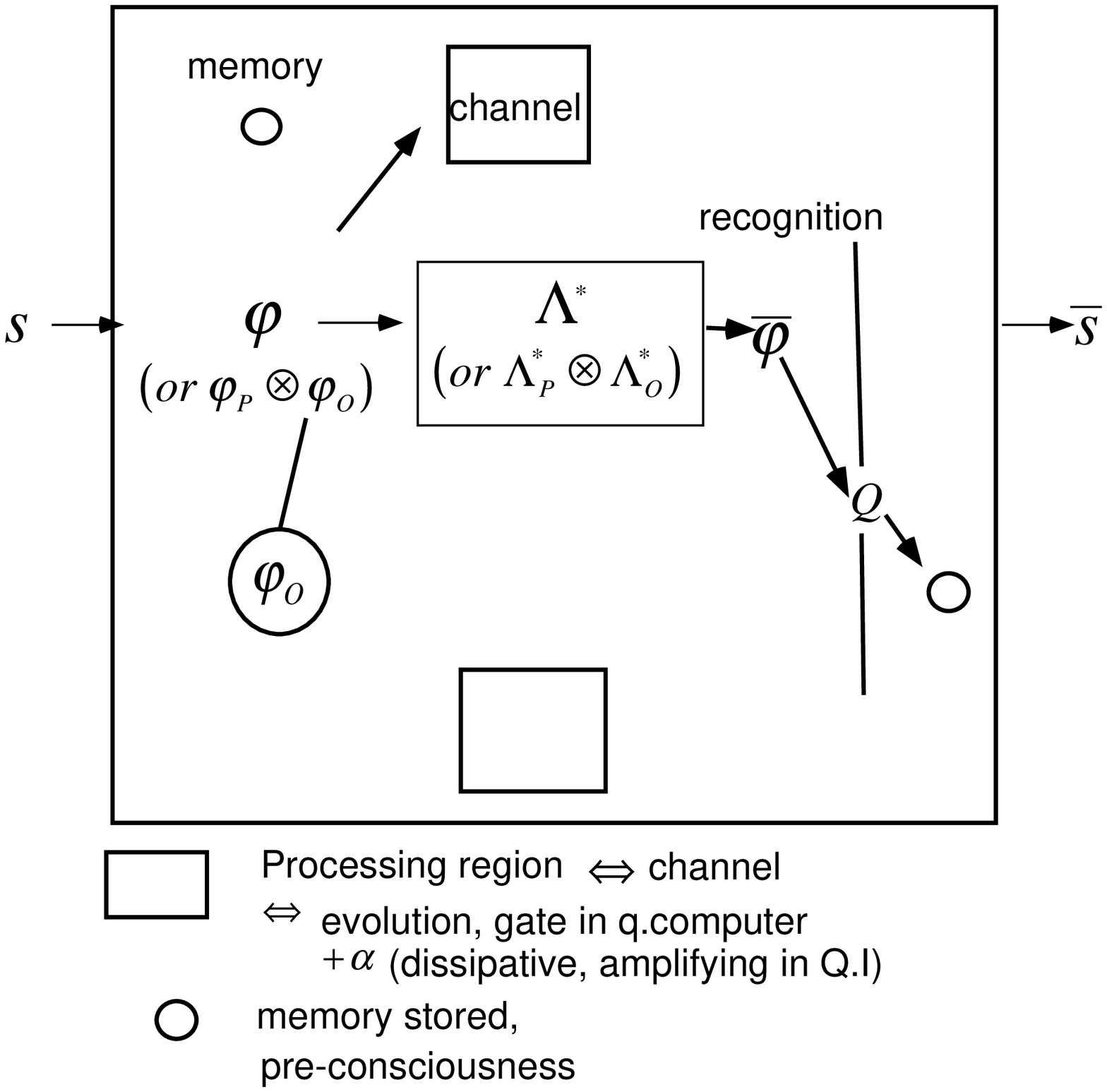}
\end{center}

Let us further assume the Hilbert space $\mathcal{H}_{O}$ is composed of two
parts, before and after recognition. For notational simplicity, we denote
the Hilbert spaces by $\mathcal{H}_{1},\mathcal{H}_{2},\mathcal{H}_{3}$
where $\mathcal{H}_{1}$ represents the processing part, $\mathcal{H}_{2}$
the memory before recognition and $\mathcal{H}_{3}=\mathcal{H}_{2}$ the
memory after recognition. Throughout this paper we will have in mind this
interpretation of the Hilbert spaces $\mathcal{H}_{j}\left( j=1,2,3\right).$ 
However, this is just an illustration of what we are going to do, and the
teleportation scheme may be applied to very different situations.

We are mainly interested in the changes of the memory after the process of
recognition. For that reason we consider channels from the set of states on $%
\mathcal{H}_{1}{\otimes }\mathcal{H}_{2}$ into $\mathcal{H}_{3}$. Main
object to be measured causing the recognition is here assumed to be a
self-adjoint operator 
\[
F=\sum_{k,l=1}^{n}z_{k,l}F_{k,l} 
\]
on $\mathcal{H}_{1}{\otimes }\mathcal{H}_{2}$ where the operators $F_{k,l}$
are orthogonal projections (alternatively, we may take $F_{k,l}$ as an
operator valued measure). The channel $\Lambda _{k,l}$ describes the state
of the memory after the process of recognition if the outcome of the
measurement according to $F$ was $z_{k,l}$ and is given by 
\[
\Lambda _{k,l}({\rho }{\otimes }{\gamma }):=\frac{\mathrm{Tr}_{1,2}(F_{k,l}{%
\otimes }{\1})({\rho }{\otimes }J{\gamma }J^{\ast }))(F_{k,l}{\otimes }%
{\1})}{\mathrm{Tr}_{1,2,3}(F_{k,l}{\otimes }{\1})({\rho }{%
\otimes }J{\gamma }J^{\ast }))(F_{k,l}{\otimes }{\1})} 
\]
where ${\rho }$ and ${\gamma }$ (denoted $\rho _{O}$ above) are the state of
the processing part and of the memory before recognition and $J$ an isometry
extending from $\mathcal{H}_{2}$ to $\mathcal{H}_{2}{\otimes }\mathcal{H}%
_{3} $ and $\1$ denotes the identical operator.  The value $\mathrm{Tr}_{1,2,3}(F_{k,l}{\otimes }{\1})({\rho }{%
\otimes \emph{J}{\gamma }\emph{J}^{\ast }})(F_{k,l}{\otimes }{\1})$
represents the probability to measure the value $z_{k,l}$. So, obviously, we
have to assume that this probability is greater than 0. The state $\Lambda
_{k,l}({\rho }{\otimes }{\gamma })$ gives the state of the memory after the
process of recognition. The elements of a basis $(b_{k})_{k=1}^{n}$ of $%
\mathcal{H}_{j}$ are interpreted as elementary signals.

In this first attempt to our model described above, there appear still a lot
of effects being non-realistic for the process of recognition. Some examples (cf.
the last section) show that with this model one can describe extreme cases
such as storing the full information or total loss of memory, but - as mentioned above - that is
still far from being a realistic description.

In this paper we restrict ourselves to finite dimensional Hilbert spaces.
Moreover, we assume equal dimension of the Hilbert spaces $\mathcal{H}%
_{j}\left( j=1,2,3\right) $. It seems that infinite dimensional schemes will
lead to more realistic models. However, this is just a first attempt to
describe the brain function. Moreover, for finite dimensional Hilbert spaces
the mathematical model becomes more transparent and one can obtain easily a
general idea of the model. To indicate obvious generalizations to more
general situations and especially to infinite dimensional Hilbert spaces we
sometimes use notions and notations from the general functional analysis. 
In a forthcoming paper we will discuss a modification of the
model using general splitting procedures on a Fock space\cite{FF1,FFL}. We hope to be able
to include more realistic effects.
\vskip0.5cm
{\em Acknowledgement:} The first two authors very much appreciate the support of the research by INTAS Project
99/00545.

\section{Basic Notions}

Let $\mathcal{H}_{1},\mathcal{H}_{2},\mathcal{H}_{3}$ be Hilbert spaces with
equal finite dimension: 
\[
\dim \mathcal{H}_{j}=n,\;\;(j\in \{1,2,3\}). 
\]
First we will represent these Hilbert spaces in a way that it seems to be
convenient for our considerations. Each of the spaces $\mathcal{H}_{1},%
\mathcal{H}_{2},\mathcal{H}_{3}$ can be identified with the space ${\Bbb{C}}%
^{n}$ of $n$-dimensional complex vectors. The space ${\Bbb{C}}^{n}$ again
may be identified with the space $\{f:G\longrightarrow \mathcal{C}\}$ of all
complex-valued function on $G:=\{1,\ldots ,n\}$. The scalar product then is
given by 
\[
\langle f,g\rangle :=\sum_{k=1}^{n}\overline{\emph{f}\left( k\right) }%
g(k)=\int \overline{\emph{f}\left( k\right) }g(k)\mu (dk) 
\]
where $\mu $ is the counting measure on $G$, i.e. $\mu =\sum_{k=1}^{n}\delta
_{k}$ with $\delta _{k}$ denoting the \textsc{Dirac} measure in $k$. So,
each of the spaces $\mathcal{H}_{j}$ can be written formally as an $L_{2}$%
-space: 
\[
\mathcal{H}_{j}=L_{2}(G,\mu ):=L_{2}(G)\;\;\;\;(j\in \{1,2,3\}). 
\]
For the tensor product one obtains 
\[
f{\otimes }g(k,l)=f(k)g(l)\;\;\;\;(f,g\in L_{2}(G),k,\in G), 
\]
and we have 
\[
\mathcal{H}_{1}{\otimes }\mathcal{H}_{2}=L_{2}(G\times G,\mu \times \mu )=%
\mathcal{H}_{2}{\otimes }\mathcal{H}_{3}. 
\]
We will abbreviate this tensor product by $L_{2}(G^{2},\mu ^{2})$ or just by 
$L_{2}(G^{2})$.

By $\mathcal{B}(\mathcal{H})$ we denote the space of all bounded linear
operators on a Hilbert space $\mathcal{H}$. In $\mathcal{B}(L_{2}(G))$ the
operator of multiplication by a function $g\in L_{2}(G)$ is given by 
\[
(\mathcal{O}\sb{g}\,f)(k)=g(k)f(k)\;\;\;\;(f\in L_{2}(G),k\in G). 
\]
Observe that for all $f,g\in L_{2}(G)$ one has 
\[
\mathcal{O}_{f}\,g=\mathcal{O}_{g}\,f,\;\;\;\;\mathcal{O}_{f}^{\ast }=%
\mathcal{O}_{\overline{f}} 
\]
and for $f\in L_{2}(G)$ with $f(k)\not=0$ for all $k\in G$ it holds $%
\mathcal{O}_{f}^{-1}=\mathcal{O}_{1/f}.$

The function $\mathbf{1}$, $\mathbf{1}(k)=1$ for all $k\in G,$ obviously
belongs to $L_{2}(G)$ and $\1=\mathcal{O}_{\mathbf{1}}$ is
the identity in $\mathcal{B}(L_{2}(G))$.

Consequently, an operator of multiplication $\mathcal{O}_f$ is unitary if
and only if $|f(k)|=1$ for all $k\in G$. \label{unitaer}

Further, we will use the mapping $J$ from $L_{2}(G)$ into $L_{2}(G^{2})$
given by 
\begin{equation}
(J\,f)(k,l)=f(k)\delta _{k,l}\;\;\;\;(f\in L_{2}(G),k,l\in G)
\label{isometry}
\end{equation}
where $\delta _{k,l}$ denotes the \textsc{Kronecker} symbol. It is immediate
to see that $J$ is an isometry. For the adjoint $J^{\ast
}:L_{2}(G^{2})\longrightarrow L_{2}(G)$ we obtain 
\begin{equation}
(J^{\ast }\Phi )(k)=\Phi (k,k)\;\;\;\;(\Phi \in L_{2}(G^{2}),\;k\in G).  \label{isometry*}
\end{equation}

Observe that $G$ equiped with the operation $\oplus :G\times
G\longrightarrow G,$ $k\oplus l:=(k+l)\mathrm{mod}~n$ \label{doplus} is a
group. The operation inverse to $\oplus $ we denote by $\ominus $. Let us
remark that $k\ominus l=k-l$ in the case $k>l$ and $k\ominus l=k-l+n$ if $%
k\leq l$. We conclude that for all $k\in G$ the operator $U_{k}\in \mathcal{B%
}(L_{2}(G))$ given by 
\begin{equation}
(U_{k}\,f)(m):=f(k\oplus m)\;\;\;\;(f\in L_{2}(G))  \label{U_k}
\end{equation}
is unitary.

Now, let $(b_k)_{k=1}^n$ be an orthonormal basis in $L_2(G)$, and denote by $%
(B_k)_{k=1}^n$ the sequence of multiplication operators corresponding to the
elements of this basis, i.e. $B_k:=\mathcal{O}_{b_k},k\in G.$

\begin{lemma}
\label{lemma1} For $k,l\in G$ we put 
\begin{equation}  \label{xikl}
\xi_{k,l}:=(B_k{\otimes} U_l)J\,\mathbf{1}.
\end{equation}
The sequence $(\xi_{k,l})_{k,l\in G}$ is an orthonormal basis in $L_2(G^2).$
\end{lemma}

\bigskip \textbf{Proof:} First observe that for all $k,l\in G$ we have 
\begin{equation}
\xi _{k,l}(m,r)=b_{k}(m)\delta _{m,r\oplus l}\qquad (m,r\in G).
\label{xikl1}
\end{equation}
So one gets 
\begin{eqnarray*}
\langle \xi _{i,j},\xi _{k,l}\rangle  &=&\int \int \overline{{\xi }_{i,j}%
\emph{(m,r)}}\cdot \xi _{k,l}(m,r)\mu ^{2}(d[m,r]) \\
&=&\int \int \overline{\emph{b}_{i}\emph{(m)}}b_{k}(m)\delta _{m,r\oplus
j}\delta _{m,r\oplus l}\mu ^{2}(d[m,r])
\end{eqnarray*}
Since $j\not=l$ implies $\delta _{m,r\oplus j}\delta _{m,r\oplus l}=0,$ the
right side will be equal to $0$ in this case. Further, observe that for all $%
l,m\in G$ there exists exactly one $r\in G$ such that $r\oplus l=m$, namely $%
r=m-l$ if $l<m$ and $r=m+n-l$ in the case $l\geq m$. So we may continue the
above chain and get for the case $j=l$ 
\begin{eqnarray*}
\langle \xi _{i,l},\xi _{k,l}\rangle  &=&\int \int \overline{\emph{b}_{i}%
\emph{(m)}}b_{k}(m)\delta _{m,r\oplus l}\mu ^{2}(d[m,r]) \\
&=&\int \overline{b_{i}\emph{(m)}}\emph{b}_{k}(m)\mu (dm)=\langle
b_{i},b_{k}\rangle 
\end{eqnarray*}
Consequently, $(\xi _{k,l})_{k,l\in G}$ is an orthonormal system in $%
L_{2}(G^{2})$, and since $\mathrm{dim}~L_{2}(G^{2})=n^{2}$ it is a basis in $%
L_{2}(G^{2}).$ $\ \ \ \square $
\vskip0.2cm

We denote by $F_{i,j}\in \mathcal{B}(L_{2}(G^{2}))$ the projection onto $\xi
_{i,j}$, i.e. 
\begin{equation}
F_{i,j}:=|\xi _{i,j}\rangle \langle \xi _{i,j}|=\langle \xi _{i,j},\cdot
\rangle \xi _{i,j}.  \label{fij}
\end{equation}

\emph{Remark: }Sometimes (especially in proofs) the 'scalar product'
notation is more convenient, but in some other cases using the 'bra-ket'
symbols the statements become more transparent. So we will use both
descriptions in the sequel.

Observe that for $\Phi \in L_{2}(G^{2})$ and $i,j\in G$ one obtains 
\begin{equation}
F_{i,j}\Phi =\xi _{i,j}\sum_{v=1}^{n}\overline{b_{i}\left( v\oplus j\right) }%
\Phi (v\oplus j,v).  \label{Fij}
\end{equation}
Indeed, we get from (\ref{xikl1}) and the definition (\ref{fij}) 
\begin{eqnarray*}
F_{i,j}\Phi &=&\langle \xi _{i,j},\Phi \rangle \xi _{i,j} \\
&=&\xi _{i,j}\int \int \overline{b_{i}\left( u\right) }\delta _{u,v\oplus
j}\Phi (u,v)\mu ^{2}(d[u,v]) \\
&=&\xi _{i,j}\int \overline{b_{i}\left( v\oplus j\right) }\Phi (v\oplus
j,v)\mu (dv).
\end{eqnarray*}
In Section \ref{generalization} we investigate concrete teleportation
channels. For this we need explicit expression for the operator $(F_{i,j}{%
\otimes }\1)(\1{\otimes }J)$. Using the definition (\ref
{isometry}) of the imbedding operator $J$ and (\ref{Fij}) we obtain for all $%
k,l,m\in G$ 
\[
((F_{i,j}{\otimes }{\1})({\1}{\otimes }J)\Phi )(k,l,m)=\xi
_{i,j}(k,l)\sum_{v=1}^{n}\overline{b_{i}\left( v\oplus j\right) }\Phi
(v\oplus j,v)\delta _{v,m} 
\]
what leads to 
\begin{equation}
((F_{i,j}{\otimes }{\1})({\1}{\otimes }J)\Phi )(k,l,m)=\xi
_{i,j}(k,l)\overline{b_{i}\left( m\oplus j\right) }\Phi (m\oplus j,m)
\label{Fij1}
\end{equation}
for all $\Phi \in L_{2}(G^{2})$ and $i,j,k,l,m\in G$.

Now, we put for $i,j\in G$ 
\begin{equation}
\hspace{-3cm}G_{i,j}:=J^{\ast }(U_{j}{\otimes }{\1})({B}_{i}^{\ast }{%
\otimes }{\1})=J^{\ast }(U_{j}B_{i}^{\ast }{\otimes }{\1})
\label{gij}
\end{equation}
where ${B}_{i}^{\ast }=\mathcal{O}_{b_{i}}^{\ast }=\mathcal{O}_{\overline{{b}%
}_{i}}.$ For $\Phi \in L_{2}(G^{2})$ and $m\in G$ we get 
\begin{eqnarray*}
(G_{i,j}\Phi )(m) &=&((U_{j}B_{i}^{\ast }{\otimes }{\1})\Phi
)(m,m)=((B_{i}^{\ast }{\otimes }{\1})\Phi )(m\oplus j,m) \\
&=&\overline{b_{i}\left( m\oplus j\right) }\Phi (m\oplus j,m).
\end{eqnarray*}
The linear operator $G_{i,j}$ maps from $L_{2}(G^{2})$ into $L_{2}(G)$ (it
is not an isometry), and we finally get for $\Phi \in L_{2}(G^{2})$ 
\begin{equation}
(F_{i,j}{\otimes }{\1})({\1}{\otimes }J)\Phi =\xi _{i,j}{\otimes 
}G_{i,j}\Phi .  \label{Fij2}
\end{equation}

\begin{example}
\label{beispfij} Consider the orthonormal basis $(b_{k})_{k=1}^{n}=(\Delta
_{k})_{k=1}^{n}$ in $L_{2}(G)$ given by $\Delta _{k}(m)=\delta _{k,m}$. From
(\ref{xikl1}) we get 
$$\xi _{i,j}(u,v)=b_{i}(u)\delta _{u,v\oplus j}=\Delta
_{i}{\otimes }\Delta _{u}(u,v\oplus j)=\Delta _{i}{\otimes }\Delta
_{i}(u,v\oplus j)=\Delta _{i}{\otimes }\Delta _{i\ominus j}(u,v),$$ 
i.e. $\xi_{i,j}=\Delta _{i}{\otimes }\Delta _{i\ominus j}$
 and we obtain for $\Phi\in L_{2}(G^{2})$ 
\[
(G_{i,j}\Phi )(m)=\overline{{b}_{i}(m\oplus j)}\Phi (m\oplus j,m)=\Delta
_{i\ominus j}(m)\Phi (i,i\ominus j). 
\]
Summarizing, in this special case we have 
\begin{equation}
(F_{i,j}{\otimes }{\1})({\1}{\otimes }J)\Phi =\Phi (i,i\ominus
j)\Delta _{i}{\otimes }\Delta _{i\ominus j}{\otimes }\Delta _{i\ominus j}.
\label{formelfij}
\end{equation}
\end{example}

\section{Entangled States}

\begin{definition}
\label{dentstate} Let ${\gamma}$ be a state on $\mathcal{H}_2=L_2(G)$ (i.e. $%
{\gamma}$ is a positive trace-class operator with $\mathrm{Tr}({\gamma})=1$%
). The state $\mathbf{e}({\gamma})$ on $L_2(G^2)=\mathcal{H}_2{\otimes} 
\mathcal{H}_3$ given by 
\begin{equation}  \label{entstate}
\mathbf{e}({\gamma})=J{\gamma} J^*
\end{equation}
where $J$ is the isometry given by (\ref{isometry})
we call the entangled state corresponding to ${\gamma}$.
\end{definition}

\begin{example}
\label{elementary} Consider the basis $(b_{k})_{k=1}^{n}=(\Delta
_{k})_{k=1}^{n}$ defined in Example \ref{beispfij}, and let ${\gamma }$ be
the pure state $|(1/\sqrt{n})\mathbf{1><(1/\sqrt{n})1|}$ (we recall that $(1/%
\sqrt{n})\mathbf{1}(k)=1/\sqrt{n}$ for all $k\in G$). For each observable $%
A\in \mathcal{B}(L_{2}(G)$ one has $\mathrm{Tr}({\gamma }A)=\frac{1}{n}%
\sum_{k=1}^{n}A\mathbf{1}(k)$. Especially, the quantum expectation of a
multiplication operator $\mathcal{O}_{f},f\in L_{2}(G)$ will be just the
arithmetic mean: 
\[
\mathrm{Tr}({\gamma }\mathcal{O}_{f}))=\frac{1}{n}\sum_{k=1}^{n}f(k). 
\]
Observe that $(1/\sqrt{n})\mathbf{1}=(1/\sqrt{n})\sum_{m=1}^{n}\Delta _{m}$
and for all $\Phi \in L_{2}(G^{2})$ it holds $\langle \Delta _{m},J^{\ast
}\Phi \rangle =\langle \Delta _{m}{\otimes }\Delta _{m},\Phi \rangle $. This
implies for $\Phi \in L_{2}(G^{2})$ and $k,l\in G$ 
\begin{eqnarray*}
(J{\gamma }J^{\ast }\Phi )(k,l) &=&({\gamma }J^{\ast }\Phi )(k)\delta _{k,l}
\\
&=&\left\langle \frac{1}{\sqrt{n}}\sum_{m=1}^{n}\Delta _{m},J^{\ast }\Phi
\right\rangle \frac{1}{\sqrt{n}}\sum_{m=1}^{n}\Delta _{m}(k)\delta _{k,l} \\
&=&\left\langle \frac{1}{\sqrt{n}}\sum_{m=1}^{n}\Delta _{m}{\otimes }\Delta
_{m},\Phi \right\rangle \frac{1}{\sqrt{n}}\sum_{m=1}^{n}\Delta _{m}{\otimes }%
\Delta _{m}(k,l)
\end{eqnarray*}
Consequently, $\mathbf{e}({\gamma })$ is the state on $L_{2}(G^{2})=\mathcal{%
H}_{2}{\otimes }\mathcal{H}_{3}$ given by 
\[
|\frac{1}{\sqrt{n}}\sum_{m=1}^{n}\Delta _{m}{\otimes }\Delta _{m}><\frac{1}{%
\sqrt{n}}\sum_{m=1}^{n}\Delta _{m}{\otimes }\Delta _{m}|. 
\]
This state $\mathbf{e}({\gamma })$ is a special representation of the
entangled state used for the elementary teleportation model \cite{AO2}.
\end{example}

Now, let ${\rho }$ and ${\gamma }$ be states on $\mathcal{H}_{1}$ resp. $%
\mathcal{H}_{2}$, the state $\mathbf{e}({\gamma })$ (usually 
denoted by $\sigma $ \cite{FO1}) will be a state on $\mathcal{H}_{2}{\otimes }\mathcal{H}%
_{3}.$ . Remember that we assumed $\mathcal{H}_{1}=\mathcal{H}_{2}=\mathcal{H%
}_{3}=L_{2}(G)$. The numbering only indicates the meaning of the states (we
recall that $\mathcal{H}_{1}$ represents the processing part, $\mathcal{H}%
_{2}$ the memory before and $\mathcal{H}_{3}$ the memory after the
recognition process.) Then ${\rho }{\otimes }\mathbf{e}({\gamma })$ is a
state on $\mathcal{H}_{1}{\otimes }\mathcal{H}_{2}{\otimes }\mathcal{H}_{3}$
and we observe immediately 
\begin{equation}
{\rho }{\otimes }\mathbf{e}({\gamma })=({\1}{\otimes }J)({\rho }{%
\otimes }{\gamma })({\1}{\otimes }J^{\ast }).  \label{zustandgross}
\end{equation}
In Section \ref{generalization} we calculate explicitly the trace of 
\begin{equation}
(F_{i,j}{\otimes }{\1})({\rho }{\otimes }\mathbf{e}({\gamma }))(F_{i,j}%
{\otimes }{\1})=(F_{i,j}{\otimes }{\1})({\1}{\otimes }J)({%
\rho }{\otimes }{\gamma })({\1}{\otimes }J^{\ast })(F_{i,j}{\otimes }%
{\1}).  \label{formel fij1}
\end{equation}
The following proposition will be very useful for this.

\begin{proposition}
\label{satz1} Let $(g_{k})_{k=1}^{n}$ and $(h_{k})_{k=1}^{n}$ be orthonormal
systems in $L_{2}(G)$ and ${\rho }$ and ${\gamma }$ states on $L_{2}(G)$
having the following representations: 
\begin{eqnarray*}
{\rho } &=&\sum_{k=1}^{n}{\alpha }_{k}|g_{k}><g_{k}|,\hspace{1cm}{\gamma }%
=\sum_{k=1}^{n}{\beta }_{k}|h_{k}><h_{k}|, \\
{\alpha }_{k} &\geq &0,{\beta }_{k}\geq 0,\sum_{k=1}^{n}{\alpha }%
_{k}=\sum_{k=1}^{n}{\beta }_{k}=1.
\end{eqnarray*}
Then for all $i,j\in G$ 
\begin{equation}
(F_{i,j}{\otimes }{\1})({\rho }{\otimes }\mathbf{e}({\gamma }))(F_{i,j}%
{\otimes }{\1})=F_{i,j}{\otimes }\sum_{k,l=1}^{n}{{\alpha }_{k}}{\beta 
}_{l}|G_{i,j}g_{k}{\otimes }h_{l}><G_{i,j}g_{k}{\otimes }h_{l}|.
\label{formelsatz1}
\end{equation}
where $G_{i,j}$ is given by (\ref{gij}).
\end{proposition}

\textbf{Proof:}  Using especially (\ref{Fij2}) we
obtain for $i,j,k,l\in G$ and $f_{1},f_{2},f_{3}\in L_{2}(G)$ 
\begin{eqnarray*}
&&(F_{i,j}{\otimes }{\1})({\1}{\otimes }J)\langle g_{k}{\otimes }%
h_{l},\cdot \rangle g_{k}{\otimes }h_{l}({\1}{\otimes }J^{\ast
})(F_{i,j}{\otimes }{\1})(f_{1}{\otimes }f_{2}{\otimes }f_{3}) \\
&=&\left\langle (F_{i,j}{\otimes }{\1})({\1}{\otimes }J)g_{k}{%
\otimes }h_{l},f_{1}{\otimes }f_{2}{\otimes }f_{3}\right\rangle (F_{i,j}{%
\otimes }{\1})({\1}{\otimes }J)g_{k}{\otimes }h_{l} \\
&=&\langle \xi _{i,j},f_{1}{\otimes }f_{2}\rangle \langle G_{i,j}g_{k}{%
\otimes }h_{l},f_{3}\rangle \xi _{i,j}{\otimes }G_{i,j}g_{k}{\otimes }h_{l}
\\
&=&F_{i,j}(f_{1}{\otimes }f_{2})\langle G_{i,j}g_{k}{\otimes }%
h_{l},f_{3}\rangle G_{i,j}g_{k}{\otimes }h_{l}
\end{eqnarray*}
Consequently, 
\begin{eqnarray*}
&&(F_{i,j}{\otimes }{\1})({\rho }{\otimes }\mathbf{e}({\gamma }%
))(F_{i,j}{\otimes }{\1}) \\
&=&(F_{i,j}{\otimes }{\1})({\1}{\otimes }J)\left[
\sum_{k,l=1}^{n}{{\alpha }_{k}}{\beta }_{l}\langle g_{k}{\otimes }%
h_{l},\cdot \rangle g_{k}{\otimes }h_{l}\right] ({\1}{\otimes }J^{\ast
})(F_{i,j}{\otimes }{\1}) \\
&=&\sum_{k,l=1}^{n}{{\alpha }_{k}}{\beta }_{l}(F_{i,j}{\otimes }{\1})(%
{\1}{\otimes }J)\langle g_{k}{\otimes }h_{l},\cdot \rangle g_{k}{%
\otimes }h_{l}({\1}{\otimes }J^{\ast })(F_{i,j}{\otimes }{\1}) \\
&=&\sum_{k,l=1}^{n}{{\alpha }_{k}}{\beta }_{l}F_{i,j}{\otimes }\langle
G_{i,j}g_{k}{\otimes }h_{l},\cdot \rangle G_{i,j}g_{k}{\otimes }h_{l} \\
&=&F_{i,j}{\otimes }\sum_{k,l=1}^{n}{{\alpha }_{k}}{\beta }_{l}\langle
G_{i,j}g_{k}{\otimes }h_{l},\cdot \rangle G_{i,j}g_{k}{\otimes }h_{l}
\end{eqnarray*}
what ends the proof.\hfill $\square $

\begin{example}
\label{beispfij2} Let us return to Example \ref{beispfij}, and suppose ${%
\rho }$ and ${\gamma }$ are given as above but with $g_{k}=h_{k}=\Delta _{k}$%
. Then 
\[
\sum_{k,l=1}^{n}{{\alpha }_{k}}{\beta }_{l}G_{i,j}\Delta _{k}{\otimes }%
\Delta _{l}=\sum_{k,l=1}^{n}{{\alpha }_{k}}{\beta }_{l}\Delta _{k}(i)\Delta
_{l}(i\ominus j)\Delta _{i\ominus j}={{\alpha }_{i}}{\beta }_{i\ominus
j}\Delta _{i\ominus j} 
\]
Consequently, 
\[
(F_{i,j}\otimes {\1})({\rho }\otimes \mathbf{e}({\gamma }%
))(F_{i,j}\otimes {\1})={{\alpha }_{i}}{\beta }_{i\ominus j}\left|
\Delta _{i}\otimes \Delta _{i\ominus j}\otimes \Delta _{i\ominus
j}\right\rangle \left\langle \Delta _{i}{\otimes }\Delta _{i\ominus j}{%
\otimes }\Delta _{i\ominus j}\right| 
\]
\end{example}

\section{Channels}

\label{channels} Denote by $\mathcal{T}$ the set of all positive trace-class
operators on $L_2(G)$ including the null operator $\mathbf{0}$, 
\[
\mathbf{0}(f)=0\hspace{3cm}(f\in L_2(G)). 
\]
We fix an operator $\tau\in\mathcal{T}$ having the representation 
\begin{equation}  \label{darst1}
\tau=\sum_{k=1}^n\gamma_k| h_k>< h_k|
\end{equation}
with $(\gamma_k)_{k\in G}\subseteq [0,\infty)$ and $(h_k)_{k\in G}$ being an orthonormal
basis in $L_2(G)$.

The linear mapping $K_\tau: \mathcal{T}\longrightarrow \mathcal{T}$ given by 
\begin{equation}  \label{ktau}
K_\tau({\rho}):=\sum_{k=1}^n\gamma_k\mathcal{O}_{ h_k}{\rho} \mathcal{O}%
_{h_k}^*\hspace{3cm}({\rho}\in\mathcal{T})
\end{equation}
depends only on the operator $\tau$ but not on its special representation.
Indeed, the following lemma holds

\begin{lemma}
Let $\tau $ have besides (\ref{darst1}) a second representaion 
\[
\tau =\sum_{k=1}^{n}\beta _{k}|g_{k}><g_{k}| 
\]
with $(\beta _{k})_{k\in G}\subseteq [0,\infty)$ and $(g_{k})_{k\in G}$ being
an orthonormal  basis in $L_{2}(G)$. For arbitrary ${\rho }\in \mathcal{T}$ , it holds 
\begin{equation}
\sum_{k=1}^{n}\gamma_{k}\mathcal{O}_{h_{k}}{\rho }\mathcal{O}%
_{h_{k}}^{\ast }=\sum_{k=1}^{n}\beta _{k}\mathcal{O}_{g_{k}}{\rho }%
\mathcal{O}_{g_{k}}^{\ast }.  \label{basiswechsel}
\end{equation}
\end{lemma}

\textbf{Proof:} It suffices to show (\ref{basiswechsel}) for ${\rho}\in%
\mathcal{T}$ of the form ${\rho}=|f><f|$ with $f\in L_2(G).$ Since 
\[
\sum_{k=1}^n\gamma_k\langle h_k,\cdot\rangle h_k=\sum_{k=1}^n\beta_k\langle
g_k,\cdot\rangle g_k 
\]
one obtains 
\begin{eqnarray*}
\sum_{k=1}^n\gamma_k\mathcal{O}_{ h_k}{\rho} \mathcal{O}_{h_k}^*&=&%
\sum_{k=1}^n\gamma_k\langle h_kf,\cdot\rangle h_kf =\sum_{k=1}^n\gamma_k%
\mathcal{O}_f\langle h_k,\cdot\rangle h_kO_f^* \\
&=&\mathcal{O}_f\sum_{k=1}^n\gamma_k\langle h_k,\cdot\rangle h_kO_f^*=%
\mathcal{O}_f\tau\mathcal{O}_f^*=\mathcal{O}_f\sum_{k=1}^n\beta_k\langle
g_k, \cdot\rangle g_kO_f^* \\
&=&\sum_{k=1}^n\beta_k\langle g_kf,\cdot\rangle g_kf=\sum_{k=1}^n\beta_k%
\mathcal{O}_{ g_k}{\rho} \mathcal{O}_{g_k}^*.
\end{eqnarray*}
$~$\hfill $\square $
\vskip0.2cm
\begin{definition}
\label{dchannel} Denote by $\mathcal{S}$ the set of all states on $L_2(G)$
and for $\tau\in\mathcal{T}$ by $\mathcal{S}_\tau$ the set of all states ${%
\rho}$ from $\mathcal{S}$ with the property that $\mathrm{Tr}K_\tau({\rho%
})$ is positive: 
\begin{equation}\label{stau}
\mathcal{S}_\tau:=\{{\rho}\in\mathcal{S}:\mathrm{Tr}K_\tau({\rho})>0\}. 
\end{equation}
For $\tau\in\mathcal{T}$ the mapping $\hat{K_\tau}:\mathcal{S}_\tau
\longrightarrow\mathcal{S}$ given by 
\begin{equation}  \label{channel}
\hat{K_\tau}({\rho}):=\frac{1}{\mathrm{Tr}K_\tau({\rho})}K_\tau({\rho})%
\hspace{2cm}({\rho}\in\mathcal{S}_\tau)
\end{equation}
is called the \textsc{channel} corresponding to $\tau$. The channel
corresponding to $\tau$ is called \textsc{unitary} if there exists an
unitary operator $U$ on $L_2(G)$ such that $\hat{K_\tau}({\rho})=U{\rho}
U^*$
\end{definition}

Observe that the channel $\hat{K_\tau}$ is in general nonlinear. However, in
Examples \ref{beisp2} and \ref{beisp3} below the channels are even unitary.

Let us make some remarks on the physical meaning of the channels ${K_\tau}$ and $\hat{K_\tau}$. The channels ${K_\tau}$ are 
mixtures of linear channels of the type
$$K^h(\rho):=\mathcal{O}_h\rho\mathcal{O}_h^*\hspace{2cm}(\rho\in\mathcal T)$$
with $h\in L_2(G),~||h||=1.$  Let us consider the more general case
$$||h||>0,~~~~|h(k)|\le 1\hspace{2cm}(k\in G).$$
We define an operator
$t_h:L_2(G)\longrightarrow L_2(\{1,2\}\times G)$
 by setting for all $f\in L_2(G)$ and $k\in G$
$$(t_h\, f)(l,k)=\left\{\begin{array}{ll}
h(k)f(k)& \mbox{~~for~~}l=1\\
&\\
\sqrt{1-|h(k)|^2}f(k)&\mbox{~~for~~}l=2.
\end{array}\right.
$$
The operator $t_h$ is an isometry from $L_2(G)$ to $L_2(\{1,2\}\times G)\cong L_2(\{1,2\})\otimes L_2(G).$ Indeed,
$$||t_h\, f||^2=\sum_{l=1}^2\sum_{k=1}^n|t_h\,f(l,k)|^2=\sum_{k=1}^n\big(|h(k)|^2+1-|h(k)|^2\big)|f(k)|^2=||f||^2.$$
Consequently, the mapping $E_h: {\mathcal B}(L_2(\{1,2\}\times G))\longrightarrow  {\mathcal B}(L_2(G))$ given by 
$$E_h(B):=t_h^*Bt_h$$
is completely positive and identity preserving. The channel $~E_h^*(\rho)=t_h\rho t_h^*~$ is the corresponding linear channel from 
the set of states on $L_2(G)$ into the set of states on $L_2(\{1,2\}\times G).$ 
The space $L_2(\{1,2\}\times G)$  has an orthogonal decomposition into $L_2(\{1\}\times G)$  and $L_2(\{2\}\times G)$  both 
being trivially isomorphic to 
$L_2(G)$. Performing a measurement according to the projection onto $L_2(\{1\}\times G)\cong L_2(G)$  given the state 
$E_h^*(\rho)$ one obtains  the state $\hat{K}^h(\rho).$  A measurement according to the projection onto $L_2(\{2\}\times G)\cong 
L_2(G)$  leads to the state $\hat{K}^{\sqrt{1-|h|^2}}(\rho)$.

 Finally, let us mention that from the statistical point of view one could get a deeper insight by considering the second quantization of 
that procedures. This means especially to replace pure states by the corresponding coherent states and the channel $E_h^*$ by the 
corresponding beam splitting \cite{FFL}.
\vskip0.5cm
\begin{example}
Assume $h\in L_2(G)$ fulfills $\inf_{k\in G}|h(k)|\ge c>0.$ Then the function $g\in L_2(G)$ given by
$$g(k):=\frac{c}{h(k)}\hspace{3cm}(k\in G)$$
fulfills the above conditions $||g||>0,~|g(k)|\le 1$ for all $k\in G$ and we obtain for all states $\rho$ on $L_2(G)$
$$\hat{K}^g(\hat{K}^h(\rho))=\rho.$$
\end{example}

\vskip0.5cm
\begin{example}
\label{beisp1} The identitiy $\tau =\1$,  (i.e. $\tau (g)=g$ for
all $g\in L_{2}(G)$) can be written in the form $\tau
=\sum_{k\in G}|g_{k}><g_{k}|$ where $(g_{k})_{k\in G}$ is an arbitrary orthonormal 
basis in $L_{2}(G)$. Now, let ${\rho }$ be an arbitrary element from $%
\mathcal{T},$ ${\rho }=\sum_{k\in G}{\alpha }_{k}|h_{k}><h_{k}|$ where $%
(h_{k})_{k\in G}$ is an orthonomal  basis in $L_{2}(G)$ and $({\alpha }_{k})_{k\in G}\subseteq [0,\infty)$. Then 
\begin{eqnarray*}
K_{\tau }({\rho }) &=&\sum_{k\in G}\mathcal{O}_{g_{k}}{\rho }\mathcal{O}%
_{g_{k}}^{\ast }=\sum_{k,l\in G}{\alpha }_{l}\langle g_{k}h_{l},\cdot
\rangle g_{k}h_{l} \\
&=&\sum_{k,l\in G}{\alpha }_{l}\mathcal{O}_{h_{l}}(\langle g_{k},\cdot
\rangle g_{k})\mathcal{O}_{h_{l}}^{\ast }=\sum_{l\in G}{\alpha }_{l}\mathcal{O}_{h_{l}}\left( \sum_{k\in G}\langle 
g_{k},\cdot \rangle g_{k}\right)
\mathcal{O}_{h_{l}}^{\ast } \\
&=&\sum_{l\in G}{\alpha }_{l}\mathcal{O}_{h_{l}}{\1}\mathcal{O}_{h_{l}}^{\ast }=\sum_{l\in G}{\alpha 
}_{l}\mathcal{O}_{h_{l}}\mathcal{O}_{h_{l}}^{\ast } \\
&=&\sum_{l\in G}{\alpha }_{l}\mathcal{O}_{|h_{l}|^{2}}
\end{eqnarray*}
Because of
$$\mathcal{O}_{|h_{l}|^{2}}=\sum_{j\in G}|h_l(j)|^2\langle\Delta_j,\cdot\rangle\Delta_j$$
the chain may be continued and we obtain
$$K_{\tau }({\rho })=\sum_{j\in G}\gamma_j\langle\Delta_j,\cdot\rangle\Delta_j$$
where 
$$\gamma_j=\sum_{l\in G}\alpha_l|h_l(j)|^2.$$
If $\rho$ belongs to $\mathcal{S}_\tau$   then $1=\sum_{j\in G}\alpha_j=\sum_{j\in G}\gamma_j$ and
$$\mathrm{Tr}(K_\tau(\rho))=\mathrm{Tr}(\rho)=1.$$
Consequently, $(\alpha_j)_{j\in G}$ is a probability distribution on $G$ and the channel
$$\hat{K_\tau}(\rho)=K_\tau(\rho)$$
transforms each state $\rho\in\mathcal{S}\tau$ into the corresponding ''classical'' state.

\end{example}
\vskip0.2cm
\begin{example}
\label{beisp2} If $\tau$ is the pure state corresponding to $(1/\sqrt{n})%
\mathbf{1}$ (cf. Example \ref{elementary}) one gets $K_\tau({\rho})=({1}/{n})\cdot {\rho}$ for all ${\rho}\in\mathcal{T}$. For 
$\rho\in\mathcal{S}\tau$ one obtains $\hat{K_\tau}({\rho})={\rho}.$ 
\end{example}
\vskip0.2cm

\begin{example}
\label{beisp3} Suppose $\tau =|b>< b|$ where $b\in
L_{2}(G) $ satisfies $|b(k)|=1/\sqrt{n}$ for all $k\in G$ (in Example \ref
{beisp2} we assumed $b(k)=1/\sqrt{n}$). Since $\mathrm{Tr}K_{\tau }({\rho }%
)=(1/n)\mathrm{Tr}(\rho)$ we obtain for all ${\rho }\in \mathcal{S}$ $\hat{K_{\tau }}({\rho 
})=B{\rho }B^{\ast }$ with $B=\sqrt{n}\mathcal{O}_{b}$. As we remarked on
page \pageref{unitaer} $\mathcal{O}_{h}$ is unitary if and only if $|h(k)|=1$
for all $k$. Consequently, the channel is unitary.
\end{example}

If $\tau$ is a mixed state $\hat{K_\tau}({\rho})$ usually will be mixed
even for pure states ${\rho}$. Below we give a simple example for this.

\begin{example}
\label{beisp4} Let $(\Delta_m)_{m\in G}$ denote as in Example \ref{beispfij}
the basis in $L_2(G)$ given by $\Delta_m(k)=\delta_{m,k}.$ Put $\tau=\frac{1%
}{2}\langle \Delta_1,\cdot\rangle\Delta_1 +\frac{1}{2}\langle
\Delta_2,\cdot\rangle\Delta_2$ and let ${\rho}$ be the pure state
corresponding to $\frac{1}{\sqrt{2}}(\Delta_1+\Delta_2).$ Then $K_\tau({%
\rho})=\frac{1}{2}\tau$ and $\hat{K_ \tau}({\rho}) =\tau.$
\end{example}

\begin{example}
\label{beisp5} Let $\tau=\langle f,\cdot\rangle f$ be a pure state, and
assume $f\in L_2(G)$ fulfils $f(k) \not=0$ for all $k$. Consequently, $%
1/f\in L_2(G)$ and for all ${\rho}\in\mathcal{S}=\mathcal{S}_\tau$ 
\[
\mathcal{O}_{1/f}\mathcal{O}_f{\rho}\mathcal{O}_f^*\mathcal{O}_{1/f}^*={%
\rho}. 
\]
This implies $K_\tau^{-1}=K_{\tilde{\tau}}$ with $\tilde{\tau}=\langle \frac{%
1}{f},\cdot\rangle \frac{1}{f}.$ Normalizing $\tilde{\tau}$ to a state one could
write alternatively $K_\tau^{-1}=\lambda^2 K_{\hat{\tau}}$, $\hat{\tau}%
=\langle g,\cdot\rangle g$ with $g=\frac {1}{\rho}\cdot\frac{1}{f}$ and $%
\lambda=||\frac{1}{f}||.$
\end{example}

\section{The State of the Memory after Recognition}

\label{generalization} Let us recall that for states ${\rho },{\gamma }$ on $%
L_{2}(G)$ and $i,j\in G$ 
\[
(F_{i,j}{\otimes }{\1})({\rho }{\otimes }\mathbf{e}({\gamma }))(F_{i,j}%
{\otimes }{\1}) 
\]
is a linear operator from $L_{2}(G^{3})$ into $L_{2}(G^{2}),$ and that (cf. (%
\ref{formel fij1})) it is equal 
\[
(F_{i,j}{\otimes }{\1})({\1}{\otimes }J)({\rho }{\otimes }{%
\gamma })({\1}{\otimes }J^{\ast })(F_{i,j}{\otimes }{\1}). 
\]
In the following we consider the family of channels $(\Lambda
_{i,j})_{i,j\in G}$ from the set of product states ${\rho }{\otimes }{\gamma 
}$ on $\mathcal{H}_{1}{\otimes }\mathcal{H}_{2}$ into the states on $%
\mathcal{H}_{3}$ given by 
\begin{equation}
\Lambda _{i,j}({\rho }{\otimes }{\gamma }):=\frac{\mathrm{Tr}_{1,2}(F_{i,j}{%
\otimes }{\1})({\rho }{\otimes }\mathbf{e}({\gamma }))(F_{i,j}{\otimes 
}{\1})}{\mathrm{Tr}_{1,2,3}(F_{i,j}{\otimes }{\1})({\rho }{%
\otimes }\mathbf{e}({\gamma }))(F_{i,j}{\otimes }{\1})}
\label{dlambdaij}
\end{equation}
where $\mathrm{Tr}_{1,2}$ resp. $\mathrm{Tr}_{1,2,3}$ denotes the partial
trace with respect to the first two components resp. the full trace with
respect to all three spaces. In the sequel we always will assume that 
\begin{equation}
\mathrm{Tr}_{1,2,3}(F_{i,j}{\otimes }{\1})({\rho }{\otimes }\mathbf{e}(%
{\gamma }))(F_{i,j}{\otimes }{\1})>0.  \label{>0}
\end{equation}
Let ${\rho }$ and ${\gamma }$ are given as in Proposition \ref{satz1}. Since 
$(\xi _{i,j})_{i,j\in G}$ is an orthonormal basis in $L_{2}(G^{2})$ (Lemma 
\ref{lemma1}) we get from Proposition \ref{satz1} 
\begin{equation}
\mathrm{Tr}_{1,2}(F_{i,j}{\otimes }{\1})({\rho }{\otimes }\mathbf{e}({%
\gamma }))(F_{i,j}{\otimes }{\1})=\sum_{k,l=1}^{n}{{\alpha }_{k}}{%
\beta }_{l}\langle G_{i,j}g_{k}{\otimes }h_{l},\cdot \rangle \ G_{i,j}g_{k}{%
\otimes }h_{l}
\end{equation}

Summarizing, we get the following representation of $\Lambda_{i,j}$:

\begin{proposition}
\label{satz5.1} Let ${\rho}$ and ${\gamma}$ be given as in Proposition \ref
{satz1}. Further, assume (\ref{>0}). Then 
\begin{equation}  \label{slambdaij}
\Lambda_{i,j}({\rho}{\otimes}{\gamma})=\frac{\sum_{k,l=1}^n{\alpha}_k{\beta}%
_l \langle G_{i,j}g_k{\otimes} h_l,\cdot\rangle\ G_{i,j}g_k{\otimes} h_l}{%
\sum_{k,l=1} ^n{\alpha}_k {\beta}_l || G_{i,j}g_k{\otimes} h_l||^2}
\end{equation}
where for $\Phi\in\mathcal{L}_2(G^2)$ 
\begin{equation}
|| G_{i,j}\Phi||^2=\sum_{m=1}^n|b_i|^2(m\oplus j)|\Phi(m\oplus j,m)|^2.
\end{equation}
\end{proposition}

\begin{example}
Let ${\rho }$ and ${\gamma }$ be pure states, ${\rho }=\langle g,\cdot
\rangle g,{\gamma }=\langle h,\cdot \rangle h.$ Then 
\begin{eqnarray*}
\mathrm{Tr}_{1,2}(F_{i,j}{\otimes }{\1})({\rho }{\otimes }\mathbf{e}({%
\gamma }))(F_{i,j}{\otimes }{\1}) &=&\langle G_{i,j}g{\otimes }h,\cdot
\rangle G_{i,j}g{\otimes }h \\
&=&\langle J^{\ast }(U_{j}\mathcal{O}_{\overline{{b_{i}}}}g{\otimes }%
h),\cdot \rangle J^{\ast }(U_{j}\mathcal{O}_{\overline{{b_{i}}}}g{\otimes }%
h).
\end{eqnarray*}
\end{example}

Fortunately, we can find expressions for the state $\Lambda _{i,j}({\rho }{%
\otimes }{\gamma })$ of the memory after the recognition process being in
many cases simpler. We can express the teleportation channel $\Lambda _{i,j}$
with the help of the channels $K_{\tau }$ we introduced in the previous
section.

\begin{proposition}\label{theorem}
Let $i,j\in G$ and let ${\rho }$ be a state from $\mathcal{S}_{|\overline{{b_{i}}}><\overline{{b_{i}}}|}$
(cf. (\ref{ktau}) and Definition \ref
{dchannel}). Further, let ${\gamma }$ be a state from $\mathcal{S}$ such
that 
\begin{equation}\label{ujk}
U_{j}K_{|\overline{{b_{i}}}><\overline{{b_{i}}}| }({\rho })U_{j}^{\ast }\in \mathcal{S}_{{\gamma }}.
\end{equation}
Then 
\begin{equation}
\Lambda _{i,j}({\rho }{\otimes }{\gamma })=\hat{K}_{{\gamma }}\circ
K^{j}\circ \hat{K}_{|\overline{{b_{i}}}><\overline{{b_{i}}}| }({\rho })  \label{lambdaij2}
\end{equation}
where $K^{j}$ denotes the unitary channel given by $K^{j}({\rho })=U_{j}{%
\rho }U_{j}^{\ast }.$
\end{proposition}

\textbf{Proof:} Let ${\rho }$ and $\gamma$ be given as in Proposition \ref{satz1}. We set 
$B_{i}=\mathcal{O}_{{b_{i}}}$. Thus $B_{i}^{\ast }=\mathcal{O}_{\overline{b_{i}}}.$ From 
$K_{|\overline{{b_{i}}}><\overline{{b_{i}}}| }({\rho })=B_{i}^{\ast }{\rho }B_{i}$ we conclude $%
K^{j}\circ K_{|\overline{{b_{i}}}><\overline{{b_{i}}}| }({\rho })=U_{j}B_{i}^{\ast }{\rho }B_{i}U_{j}^{\ast }$
what leads to

\begin{eqnarray*}
K_{{\gamma }}\circ K^{j}\circ K_{|\overline{{b_{i}}}><\overline{{b_{i}}}| }({\rho }) &=&\sum_{l=1}^{n}{\beta }%
_{l}\mathcal{O}_{h_{l}}U_{j}B_{i}^{\ast }{\rho }B_{i}U_{j}^{\ast }\mathcal{O}%
_{h_{l}}^{\ast } \\
&=&\sum_{k,l=1}^{n}{\alpha }_{k}{\beta }_{l}\mathcal{O}_{h_{l}}U_{j}B_{i}^{%
\ast }\left( \langle g_{k},\cdot \rangle g_{k}\right) B_{i}U_{j}^{\ast }%
\mathcal{O}_{h_{l}}^{\ast } \\
&=&\sum_{k,l=1}^{n}{\alpha }_{k}{\beta }_{l}\langle h_{l}U_{j}B_{i}^{\ast
}g_{k},\cdot \rangle h_{l}U_{j}B_{i}^{\ast }g_{k} \\
&=&\sum_{k,l=1}^{n}{\alpha }_{k}{\beta }_{l}\langle J^{\ast
}(U_{j}B_{i}^{\ast }{\otimes }{\1})g_{k}{\otimes }h_{l},\cdot \rangle
J^{\ast }(U_{j}B_{i}^{\ast }{\otimes }{\1})g_{k}{\otimes }h_{l} \\
&=&\sum_{k,l=1}^{n}{\alpha }_{k}{\beta }_{l}\langle G_{i,j}g_{k}{\otimes }%
h_{l},\cdot \rangle \ G_{i,j}g_{k}{\otimes }h_{l}
\end{eqnarray*}
Finally, from 
\[
\hat{K}_{{\gamma }}\circ K^{j}\circ K_{|\overline{{b_{i}}}><\overline{{b_{i}}}| }({\rho })=\hat{K}_{{\gamma }%
}\circ K^{j}\circ \hat{K}_{|\overline{{b_{i}}}><\overline{{b_{i}}}| }({\rho })
\]
we obtain (\ref{lambdaij2}). $\ \ \ \square $

\bigskip

\bigskip\ In the following let us comment the results and give some
examples. Let ${\rho }$ be an arbitrary state of the processing part (the
brain), and assume the measurement of the incoming signal leads to the value 
$z_{i,j}$. Then the input in the memory being in the state ${\gamma }$ will
be 
\[
C\cdot U_{j}B_{i}^{\ast }{\rho }B_{i}U_{j}^{\ast } 
\]
where $C$ is the normalizing constant. After the recognition process the
brain will be in the state 
\begin{equation}\label{endformel}
\tilde{C}\cdot K_{\gamma }\circ (U_{j}B_{i}^{\ast }{\rho }B_{i}U_{j}^{\ast
}) 
\end{equation}
where $\tilde{C}$ is again the normalizing constant. 

\begin{example}
Let us consider  the extreme cases that either the processing part  or the memory is in the trivial state 
\begin{equation}\label{varkappa}
\varkappa:=|\frac{1}{\sqrt{n}}\mathbf{1}><\frac{1}{\sqrt{n}}\mathbf{1}|
\end{equation}
 (cf. Example \ref{elementary}). This state has no experience, no special knowledge, there will be no selection of  incoming 
information. It is easy to check that for all $\mu\in{\mathcal S}$ it holds
\begin{equation}\label{kappamu}
\hat{K}_\varkappa(\mu)=\mu
\end{equation}

(observe that ${K}_\varkappa(\mu)=({1}/n)\cdot\mu).$
On the other hand, for all states $\mu$  the relation
$$\hat{K}_\mu(\varkappa)=\mu $$
is true.
\vskip0.2cm
Now, we consider the case  of the memory being  in the state $\gamma=\varkappa$. We obtain  from (\ref{kappamu})
and from (\ref{endformel}) 
for all $i,j\in G$ and
$\rho\in{\mathcal S}_{|\overline{{b_{i}}}><\overline{{b_{i}}}| }$
\[
\Lambda _{i,j}({\rho }{\otimes }{\gamma })=\tilde{C}\cdot U_{j}B_{i}^{\ast }{\rho }B_{i}U_{j}^{\ast }.
\]
The memory will store exactly what comes in (the system is able to learn everything - cf. also \cite{FO1,FO2}).
Since $U_j$ and $\sqrt{n}B_i$ are unitary operators ($\tilde{C}=n$)   we see that for all $i,j$ there exists (in the language of teleportation procedures) a unitary key $V_{i,j}$ to recover $\rho$, i.~e. 
$\Lambda _{i,j}({\rho }{\otimes }{\gamma })=V_{i,j}\rho V_{i,j}^*.$
\vskip0.3cm

Now, let the processing part (the brain) be in the state $\rho=\varkappa$  defined by (\ref{varkappa}).  Then 
$$K^j\circ\hat{K}_{|\overline{{b_{i}}}><\overline{{b_{i}}}| }(\rho)=|U_j\overline{{b_{i}}}><U_j\overline{{b_{i}}}| .$$
For all states of the brain $\gamma$ such that $|U_j\overline{{b_{i}}}><U_j\overline{{b_{i}}}|\in{\mathcal S}_\gamma$ we 
obtain
$$\Lambda _{i,j}({\rho }{\otimes }{\gamma })=\hat{K}_\gamma(|U_j\overline{{b_{i}}}><U_j\overline{{b_{i}}}|).$$
So (as one could expect) the final state in the memory depends only on the measured value $z_{i,j}$ and the state of the memory (before recognition).
\end{example}
\begin{example}\label{beisp19}
Let $(b_j)_{j\in G}$ be an orthonormal basis fulfilling
$$|b_j(l)|^2=\frac{1}{n}\hspace{3cm} (j,l,\in G).$$
For the pure state $\rho =\frac{1}{n}\1$ we obtain for all $i\in G$
$$K_{|\overline{{b_{i}}}><\overline{{b_{i}}}| }(\rho)=\frac{1}{n}{\mathcal O}_{|b_i|^2}=\frac{1}{n^2}\sum_{k\in 
G}|\Delta_k><\Delta_k|=\frac{1}{n^2}\1.$$
Consequently, for $j\in G$ 
$$K^j\circ K_{|\overline{{b_{i}}}><\overline{{b_{i}}}| }(\rho)=\frac{1}{n^2}\1.$$
For each state $\gamma$ (we use  notation as in Proposition \ref{satz1})  and all $i,j\in G$ we finally get
\begin{eqnarray*}
\Lambda_{i,j}(\rho\otimes \gamma)&=&\sum_{k\in G}\frac{\beta_k}{ n}{\mathcal O}_{|h_k|^2}
=\sum_{l\in G}\gamma_l|\Delta_l><\Delta_l|
\end{eqnarray*}
with
$$\gamma_l =\sum_{k\in G}\frac{\beta_k|h_k|^2(l)}{n}.$$
So we obtain a classical state with  probability distribution
$(\gamma_k)_{k\in G}.$
\end{example}
\vskip0.5cm
\begin{example}\label{beisp20}
 Take $(b_j)_{j\in G}$ as in Example \ref{beisp19} above, but now suppose $\gamma=\frac{1}{n}\1$.
Let $\rho$ be given as in Proposition \ref{satz1}. For all $i,j\in G$ we get
\begin{equation}\label{20}
K^j\circ K_{|\overline{{b_{i}}}><\overline{{b_{i}}}| }(\rho)=U_jB_i^*\rho B_iU_j^*=
\sum_{k\in G}\alpha_k \big|U_jB_i^*g_k\big>\big<U_jB_i^*g_k\big|.
\end{equation}
Observe that for all $i,j$ the  sequence $(U_jB_i^*g_k)_{k\in G}$ is an orthogonal system. Indeed, 
$$\langle U_jB_i^*g_k,U_jB_i^*g_l\rangle=\frac{1}{n}\langle U_jg_k,U_jg_l\rangle =\frac{1}{n}<g_k,g_l>.$$
Consequently, as  in Example \ref{beisp1} and above we conclude
\begin{eqnarray*}
K_\gamma\circ K^j\circ K_{|\overline{{b_{i}}}><\overline{{b_{i}}}| }(\rho)&=&
\sum_{k\in G}\frac{\alpha_k}{n} {\mathcal O}_{|U_jB_i^*g_k|^2}\\
&=&\sum_{k\in G}\frac{\alpha_k}{n^2}{\mathcal O}_{|U_jg_k|^2}\\
&=&U_j\sum_{k\in G}\frac{\alpha_k}{n^2}{\mathcal O}_{|g_k|^2}U_j^*\\
&=&U_j\sum_{k\in G}\gamma_k|\Delta_k><\Delta_k|U_j^*\\
&=&\sum_{k\in G}\gamma_k|\Delta_{k\ominus j}><\Delta_{k\ominus j}|
\end{eqnarray*}
with
$$\gamma_k=\sum_{l\in G}\frac{\alpha_l}{n^2}|g_l|^2(k).$$
Finally we thus get 
$$\Lambda_{i,j}(\rho\otimes \gamma)=
\sum_{k\in G}\tilde{\gamma_k}|\Delta_{k\ominus j}><\Delta_{k\ominus j}|$$
with 
$$\tilde{\gamma_l} =\sum_{k\in G}\alpha_k|h_k|^2(l).$$
We see that $\Lambda_{i,j}$ does not depend on $i$. Especially, there do not exist unitary keys.
\end{example} 

\begin{example} We take again  $(b_j)_{j\in G}$ as in Examples \ref{beisp19} and \ref{beisp20}. Let us further assume  that $\gamma=|h><h|$ is a pure state satisfying
$|h(j)|^2=\frac{1}{n}$ for all $j\in G$. 
Using (\ref{20}) we get for arbitrary $\rho$
\begin{eqnarray*}
K_\gamma\circ K^j\circ K_{|\overline{{b_{i}}}><\overline{{b_{i}}}| }(\rho)&=&
{\mathcal O}_h U_jB_i^*\rho B_iU_j^*{\mathcal O}_h^* 
\end{eqnarray*}
Observe that $\sqrt{n}{\mathcal O}_h$ and $\sqrt{n}B_i$ are unitaries. Consequently, we get 
$$\Lambda_{i,j}(\rho\otimes \gamma)=V_{i,j}\rho V_{i,j}^*$$
with the unitary key $V_{i,j}=n{\mathcal O}_h U_jB_i^*.$
\end{example}

\bigskip

The choice of the basis $(b_{k})_{k\in G}$ is very important in this model.
Because of the specially chosen projection operators $F_{i,j}$ these are the
only elementary signals that can be measured. Let us consider the case that
the selected basis $(b_{k})_{k=1}^{n}$ is given by $(\Delta _{k})_{k=1}^{n}$.
In this case we get an especially simple (but also trivial) output. Let us
remark that for all $r,k,l\in G$ such that $k\not=l$ it holds $\Delta
_{k}(r)\Delta _{l}(r)=0$. Thus the elements of the basis fulfil a condition
much more stringent than just being orthogonal.

\begin{example}
In Example \ref{beispfij} we obtain for ${\rho }=\sum_{k=1}^{n}{\alpha }%
_{k}|\Delta _{k}><\Delta _{k}|,{\gamma }=\sum_{k=1}^{n}{\beta }_{k}|\Delta
_{k}><\Delta _{k}|$ 
\[
\mathrm{Tr}_{1,2}(F_{i,j}{\otimes }{\1})({\rho }{\otimes }\mathbf{e}({%
\gamma }))(F_{i,j}{\otimes }{\1})={{\alpha }_{i}}{\beta }_{i\ominus
j}|\Delta _{i\ominus j}><\Delta _{i\ominus j}| 
\]
and if ${{\alpha }_{i}}>0,{\beta }_{i\ominus j}>0$ 
\[
\Lambda _{i,j}({\rho }{\otimes }{\gamma })=|\Delta _{i\ominus j}><\Delta
_{i\ominus j}|. 
\]
So if ${{\alpha }_{i}}>0,{\beta }_{i\ominus j}>0$ the state after
recognition depends only on the measured value $i\ominus j$. If ${\rho }$
resp. ${\gamma }$ cannot occur in the state $\Delta _{i}$ resp. $\Delta
_{i\ominus j}$ no information about the input can be stored in the memory.
\end{example}

\begin{example}
Let ${\rho }=\sum_{k=1}^{n}{\alpha }_{k}|g_{k}><g_{k}|,~{\gamma }=\sum_{k=1}^{n}{\beta }_{k}|h_{k}><h_{k}|$ be arbitrary 
states. What will be
the state of the memory after recognition if only these elementary signals $\Delta_k$
can be measured? Will these elementary signals also damage the states ${\rho 
}$ and ${\gamma }$ in such a way that only the information whether ${{\alpha 
}_{i}}>0$and ${\beta }_{i\ominus j}>0$ plays a role? Simple calculations as
above lead to the following form (in what follows we omit normalizing
constants): 
\[
\mathrm{Tr}_{1,2}(F_{i,j}{\otimes }{\1})({\rho }{\otimes }\mathbf{e}({%
\gamma }))(F_{i,j}{\otimes }{\1})=\sum_{k,l=1}^{n}{\alpha }_{k}{\beta }%
_{l}|A_{k,l}><A_{k,l}| 
\]
where 
\[
A_{k,l}(r)=\Delta _{i\ominus j}(r)g_{k}(r\oplus j)h_{l}(r)\hspace{3cm}(r\in
G). 
\]
Now, for each $k\in G$ there exist sequences $(\mu _{k}^{s})_{s=1}^{n},(\nu
_{k}^{s})_{s=1}^{n}$ such that $g_{k}=\sum_{s=1}^{n}\mu _{k}^{s}\Delta _{s}$
and $h_{k}=\sum_{s=1}^{n}\nu _{k}^{s}\Delta _{s}$. This implies $A_{k,l}=\mu
_{k}^{i}\nu _{l}^{i\ominus j}\Delta _{i\ominus j}.$ Consequently, 
\begin{eqnarray*}
\mathrm{Tr}_{1,2}(F_{i,j}{\otimes }{\1})({\rho }{\otimes }\mathbf{e}({%
\gamma }))(F_{i,j}{\otimes }{\1}) &=&\sum_{k,l=1}^{n}{\alpha }_{k}{%
\beta }_{l}|\mu _{k}^{i}\nu _{l}^{i\ominus j}|^{2}|\Delta _{i\ominus
j}><\Delta _{i\ominus j}| \\
&=&C\cdot |\Delta _{i\ominus j}><\Delta _{i\ominus j}|.
\end{eqnarray*}
The state $\sigma =\Lambda _{i,j}({\rho }{\otimes }{\gamma })$ after
recognition will be the same as in the above example. We see that measuring $
z_{i,j}$ the state $\sigma $ will be able to store in his memory at  most  the
signal $\Delta _{i\ominus j}.$  And this can be done only if  there exists at least one pair $(k,~l) $ such that 
${\alpha }_{k}{\beta }_{l}|\mu _{k}^{i}\nu _{l}^{i\ominus j}|^{2}>0$..
\end{example}

\bigskip

\textbf{Concluding remarks:} The aim of the paper was to touch the problem
of finding simplified models for the recognition process. We were interested in  how the input signal arriving at the brain
is entangled (connected) to the memory already stored and the consciousness that
existed in the brain, and how  a part of the signal will be finally stored as a
memory. Just to achieve simple explicit expressions we illustrated the model
on the most simple sequence of signals $(\Delta _{k})_{k=1}^{n}$. It is
clear that this example is just for illustration and can not serve for
describing realistic aspects of recognition. Choosing a more complex basis
one obtains expressions depending heavily on the states ${\rho }$ and ${%
\gamma }$. Though the above presented model is only a first attempt it shows
that there are possibilities to model the process of recognition. To get
closer to realistic models we will try to refine the above models by

\begin{itemize}
\item[-]  passing over to infinite Hilbert spaces,

\item[-] replacing pure states by coherent states on the Fock space,

\item[-]  considering different Hilbert spaces $\mathcal{H}_1,\mathcal{H}_2$
and $\mathcal{H}_3$,

\item[-]  making more complex measurements than simple one-dimensional
projections $F_{i,j}$,

\item[-]  replacing the trivial entanglement $J$ by a more complex one based on beam splitting procedures, and
finally

\item[-]  adding  an entanglement
between the states ${\rho}$ and ${\gamma}$ on $\mathcal{H}_1$ and $\mathcal{H%
}_2$.
\end{itemize}

\bibliographystyle{plain}

\end{document}